\documentclass[11pt,a4]{article}

\usepackage{graphicx}

\begin{document}

\vspace*{1cm}
\centerline{\LARGE \bf Knots and Links in Physical Systems
\footnote{Presented by TWK at ``Knots and soft-matter physics'', YITP, Kyoto, August, 2008.}} %

\begin{flushright}
{\large Roman V. Buniy~\footnote{E-mail: roman.buniy@gmail.com}, \\
Physics Department, Indiana University, Bloomington, IN 47405, USA \\
Martha J. Holmes~\footnote{E-mail: martha.j.holmes@vanderbilt.edu} and
Thomas W. Kephart~\footnote{E-mail: tom.kephart@gmail.com} \\
Department of Physics and Astronomy, Vanderbilt University, Nashville,
TN 37235, USA}
\end{flushright}

\vspace{5mm}

Abstract: Few physical systems with topologies more complicated than
simple gaussian linking have been explored in detail. Here we focus on
examples with higher topologies in non-relativistic quantum mechanics
and in QCD.

\section{Generalized Aharonov-Bohm and Josephson effects}

Topology has played an important role in physics in recent years, but
unlike in biology, most work done so far has involved the simplest
nontrivial topology---gaussian linking. Here we discuss more complex
topologies and their implications for physical systems.
 
Let us start by recalling two prominent examples of the use of
gaussian linking in non-relativistic quantum mechanics, the magnetic
Aharonov-Bohm effect~\cite{Aharonov:1959fk} and the Josephson
effect~\cite{josephson}, see also Ref.~\cite{FeynmanLectures}. We can
generalize both these systems to higher order linking and/or
knotting~\cite{Rolfsen}.
 
\begin{figure}[h] 
  \includegraphics[width=6cm]{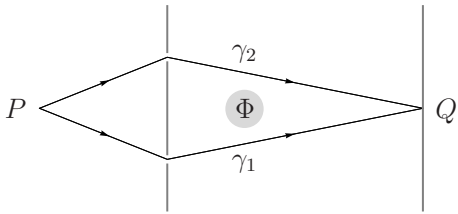}
  \caption{\label{figure-AB-apparatus} A plane projection of the standard
    magnetic Aharonov-Bohm effect apparatus.}
\end{figure}

The magnetic Aharonov-Bohm effect, see Fig.~\ref{figure-AB-apparatus},
results when a charged particle travels around a closed path in a
region of vanishing magnetic field but nonvanishing vector potential.
The wave function of the particle is affected by the vector potential
and a vector potential dependent interference pattern proportional to
magnetic flux occurs at a detection screen. The conclusion one draws
is that the vector potential is more fundamental than the magnetic
field. The definitive experiments were done here in
Japan~\cite{Tomoyama}.

\begin{figure}[h]
  \includegraphics[width=6cm]{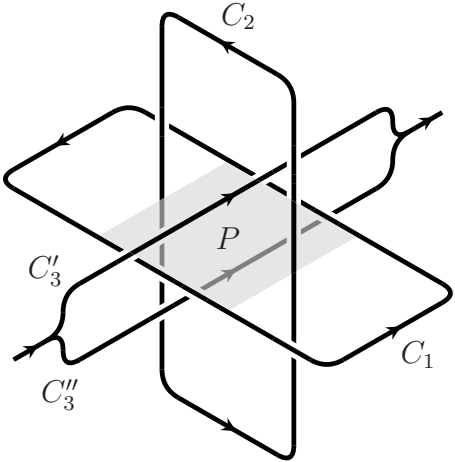}
  \caption{\label{figure-experiment} Schematic of a Borromean ring
    arrangement to detect the second order phase.}
\end{figure}

Shown in Fig.~\ref{figure-experiment} is a schematic of a Borromean
ring arrangement to detect the second order phase $\phi_{12}$, where
$C_1$ and $C_2$ are magnetic solenoids carrying flux $\Phi_1$ and
$\Phi_2$, and $C_3'$ and $C_3''$ correspond to two topologically
distinct semi-classical paths and are parts of the closed path $C_3$
for the wave function of a charged particle starting from the source
and ending at the screen \cite{Buniy:2006tq}. To prevent gaussian
linking of the wave function with the solenoids there is a rectangular
plate in the plane of $C_1$.  We predict interference proportional to
$ (e/\hbar c)^2\Phi_1\Phi_2$. For the case of knotting see
Ref.~\cite{Buniy:2008cg}.

\begin{figure} 
  \includegraphics[width=6.5cm]{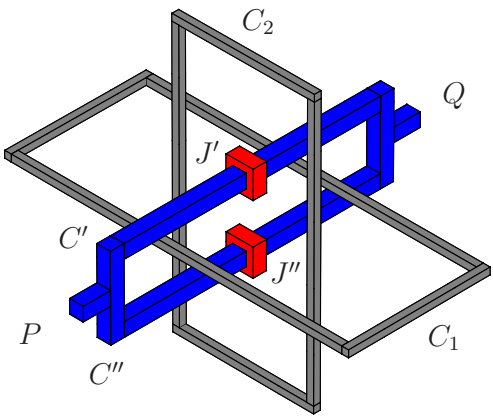}
  \caption{\label{figure-GenJJ-apparatus} A schematic of a generalized
    magnetic Josephson effect apparatus. $J'$ and $J'$ are Josephson
    junctions. $C_1$ and $C_2$ are solenoids.}
\end{figure}

A similar generalization of the Josephson experiment to higher order
linking, is possible~\cite{Buniy:2008ch}, see
Fig.~\ref{figure-GenJJ-apparatus}. There one predicts the maximum
current flowing through the superconductor from $P$ to $Q$ to be
\begin{equation}
  J_\textrm{max} = J_0 \left| \cos\left( \frac{\pi\Phi_1
    \Phi_2}{\Phi_0}\right) \right|,
\end{equation}
where $\Phi_0$ is the fluxoid.

\section{Tightly Knotted QCD Flux Tubes as Glueballs}

\begin{figure} 
  \includegraphics[width=15cm]{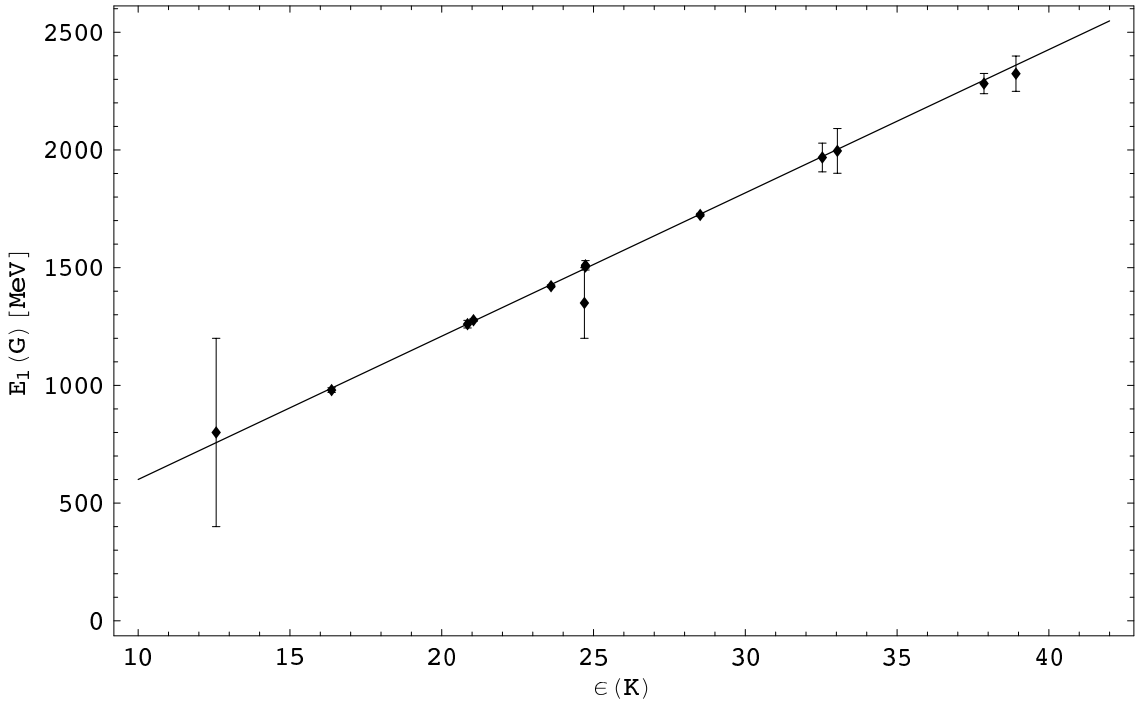}
  \caption{\label{tier1} Preliminary fit of $f_0$ states to the tight
  knot/link spectrum.}
\end{figure}

The study of bosonic exchange particles in hadronic physics have a
long illustrious history going back to the work of
Yukawa~\cite{Yukawa:1935xg}. The quark model is sufficient to describe
most of the spectrum of hadronic bound states, but after filling the
multiplets, a number of states remain and it has been suggested that
at least one of these states is a glueball---states with no valance
quarks. Two of us have suggested that the glueball spectrum of QCD is
a result of tight knots and links of quantized chromo-electric
flux~\cite{Buniy:2002yx}. (The study of tight knots started in
biology, see Ref.~\cite{ASetal}.) This provides an infinite spectrum,
up to stability of new hadronic states, and predicts their
energies. Identifying knot lengths with particle energies means the
glueball spectrum is the same as the tight knot spectrum up to an
overall scaling parameter. A preliminary fit matching the most recent
knot/link lengths~\cite{ACPR} with the presumed glueball
states~\cite{Amsler:2008zz} of zero angular momentum, the $f_0$
states, is shown in Fig.~\ref{tier1}, see Ref.~\cite{BHK}.

Other systems of tightly knotted fluxes or other types of
matter/energy should all generate universal behavior, with spectra all
corresponding to the length of knots and links up to a single overall
scaling parameter~\cite{Buniy:2002yx}.

In the present model, glueball candidates of non-zero angular
momentum, called $f_J$ states, correspond to spinning knots and
links. To calculate rotational energies we need the moment of inertia
tensor for each tight knot and link. We can get exact results for
links with planar components. E.g., in its center of mass frame, the
moment of inertia tensor for a Hopf link of uniform density is
\begin{eqnarray}{I_{\textrm{Hopf}}} = 
\left( \begin{array}{ccc} 21 & 0 & 0 \\ 0 & \frac{75}{2} & 0 \\ 0 & 0
& \frac{75}{2}
\end{array} \right)\pi^{2} \rho \;a^5,
\label{eq11}
\end{eqnarray}
where one torus is in the $xz$-plane and the other is in the
$xz$-plane. Note this inertia tensor corresponds to a prolate spheroid
as do other straight chains of links with an even number of elements.
Odd straight chains do not have so much symmetry. Moment of inertia
tensors of other knots and links can be calculated by Monte Carlo
methods~\cite{JCetal}. For further discussion and a detailed analysis
of both the zero and non-zero angular momentum $f_J$ states including
calculations of moment of inertia tensor see Ref.~\cite{MJHthesis}.

\section*{Acknowledgments}

TWK thanks the Yukawa Institute Theoretical Physics and the Vanderbilt
College of A\&S for travel support. The work of RVB was supported by
DOE grant number DE-FG02-91ER40661 and that of MJH and TWK by DOE
grant number DE-FG05-85ER40226.

\end{document}